\documentclass{article}
\usepackage{graphicx}  
\usepackage{amsmath}   
\usepackage[compress]{cite}
\usepackage{amssymb}   
\usepackage{bm} 
\usepackage{dcolumn}
\usepackage{color}
\usepackage{mathrsfs}
\usepackage{amsfonts}
\usepackage{varioref}
\RequirePackage[colorlinks,citecolor=blue,urlcolor=magenta,linkcolor=blue]{hyperref}
\def\EH{Einstein-Hilbert }

\def\gr{general relativity}

\def\KR{Kalb-Ramond }

\def\EGB{Einstein-Gauss-Bonnet }
\def\KK{Kaluza-Klein }
\labelformat{section}{Section #1} 
\labelformat{subsection}{Section #1} 
\labelformat{subsubsection}{Section #1}
\labelformat{subsubsubsection}{Section #1}
\labelformat{equation}{Eq.~(#1)} 
\labelformat{figure}{Fig.~#1} 
\labelformat{subfigure}{Fig.~\thefigure#1} 
\labelformat{table}{Tab.~#1} 
\labelformat{appendix}{Appendix #1}
\title{Packing extra mass in compact stellar structures: An interplay between Kalb-Ramond field and extra dimensions}
\author{Sumanta Chakraborty\footnote{sumantac.physics@gmail.com}
and Soumitra SenGupta\footnote{tpssg@iacs.res.in}\\
{\small{Department of Theoretical Physics,
Indian Association for the Cultivation of Science, Kolkata-700032, India}}}
\begin{document}

\maketitle
\begin{abstract}
We have derived the Buchdahl's limit for a relativistic star in presence of the Kalb-Ramond field in four as well as in higher dimensions. It turns out that the Buchdahl's limit gets severely affected by the inclusion of the Kalb-Ramond field. In particular, the Kalb-Ramond field in four spacetime dimensions enables one to pack extra mass in any compact stellar structure of a given radius. On the other hand, a completely opposite picture emerges if the Kalb-Ramond field exists in higher dimensions, where the mass content of a compact star is smaller compared to that in general relativity. Implications are discussed.
\end{abstract}

\section{Introduction}\label{Buch_Tor_Intro}

There have been considerable interest in the compactness limit of any stellar structure, which originally initiated from the seminal work of Buchdahl, who showed that under reasonable assumptions the minimum radius of a star has to be greater than (9/8) of its Schwarzschild radius \cite{Buchdahl:1959zz,Wald,gravitation}. These assumptions involved the density of the star to be decreasing outwards and the interior solution being matched to the vacuum exterior one, which by uniqueness theorems of \gr\ is the Schwarzschild solution. This raises an intriguing question, how is the above limit modified if one considers a theory of gravity different from \gr\ or if one introduces some additional matter fields. Several such ideas have been explored quiet extensively in recent times, which include --- (a) inclusion of cosmological constant 
\cite{Mak:2001gg,Andreasson:2012dj,Stuchlik:2008xe,Zarro:2009gd}, (b) effects due to presence of extra dimensions 
\cite{PoncedeLeon:2000pj,Germani:2001du,Garcia-Aspeitia:2014pna,harko-mak}, (c) effect of scalar tensor theories
\cite{Singh:1983qp,Yokoi:1972hi,Burikham:2016cwz,Tsuneishi:2005um,Horbatsch:2010hj,Pani:2011xm,Horbatsch:2012hla,Tsuchida:1998jw} and (d) dependence of Buchdahl's limit on higher curvature gravity models 
\cite{Goswami:2015dma,Das:2014rca,Wright:2015yda,Dadhich:2010qh,Dadhich:2016fku,Hendi:2017ibm,Dadhich:2016wtb,
Molina:2016xeu,Burikham:2016cwz,Banerjee:2017mbv}, such as, \EGB gravity, $f(R)$ gravity \cite{Nojiri:2010wj,Sotiriou:2008rp,Nojiri:2003ft,DeFelice:2010aj,Chakraborty:2016ydo,Chakraborty:2016gpg}, pure Lovelock theories \cite{Padmanabhan:2013xyr,Chakraborty:2015wma,Dadhich:2008df,Kastor:2012se,
Chakraborty:2017bcg} etc. In some of these cases one had to impose some additional physically motivated assumptions, e.g., imposition of dominant energy condition, sub-luminal propagation of sound etc. All in all, the ultimate limit on stellar structures is of fundamental importance for not only understanding the differences between various gravity theories but also to differentiate how additional matter fields behave under self-gravity \cite{Breu:2016ufb,Fujisawa:2015nda,Barraco:2002ds,Karageorgis:2007cy,
Andreasson:2007ck}. 

Having set the stage, let us consider the situation we will be interested in. Rather than working with modified gravity theories, we will modify Einstein's field equations by introducing additional matter fields. In particular we will mainly be concerned about the effects of the Kalb-Ramond field in the formation of any stellar structure and possible modification of the Buchdahl's limit thereof. Kalb-Ramond field arises naturally in the context of field theory (and also as a closed string mode in string theory) and is sort of a generalization of electrodynamics \cite{Green:1987mn,Kalb:1974yc,Kar:2001eb,Maity:2004he,Majumdar:1999jd,Mukhopadhyaya:2004cc,Kar:2002xa,Chakraborty:2016lxo}. The gauge vector field in electrodynamics gets replaced by a second rank antisymmetric tensor field with a corresponding third rank antisymmetric field strength. Interestingly, the corresponding third rank antisymmetric tensor field appearing as field strength of the Kalb-Ramond field has conceptual as well as mathematical 
similarity with the appearance of spacetime torsion \cite{Hehl:1976kj,deSabbata:1994wi,Letelier:1995ze,Capozziello:2001mq,Mukhopadhyaya:2002jn,SenGupta:2001cs,SenGupta:2001yj,
Lue:1998mq,Sur:2016bwu}. In particular if one assumes spacetime torsion to be antisymmetric in all the three indices\footnote{In general, spacetime torsion appears through the definition of covariant derivative: $\nabla_{i}V^{j}=\partial _{i}V^{j}+A^{j}_{ik}V^{k}$, with connection being written as $A^{a}_{bc}=\Gamma ^{a}_{bc}+T^{a}_{bc}$. Here $\Gamma ^{a}_{bc}$ is the standard Christoffel connection, symmetric in $(b,c)$, while $T^{a}_{bc}$ is the torsion tensor, antisymmetric in $(b,c)$.}, then the decomposition of the Ricci scalar into parts dependent and independent of torsion leads to \EH action with an additional term coinciding with the action for \KR field coupled to gravity. Thus in this sense, whether we work with \KR field or completely antisymmetric spacetime torsion the physics would remain unchanged. However for concreteness we will explicitly work with \KR field 
while keeping the analogy with spacetime torsion in the backdrop. Thus given the action for the \KR field in presence of gravity one can immediately obtain the corresponding modified Einstein's equations. This will result into different exterior as well as interior solution, in particular the exterior solution no longer represents a vacuum spacetime. This will inevitably results into modifications on the limits of stellar structure and hence the Buchdahl's limit would certainly be different from that in \gr. There is another way a similar modification can be brought about, which corresponds to introduction of extra dimensions. Even though in the standard picture the normal matter fields, e.g., electromagnetic field does not propagate in the extra dimensions\footnote{This is due to the fact that all such matter fields originate from open string modes and hence has access to four-dimensional spacetime only as their end points are attached to it.}, the \KR on the other hand, being a closed string mode alike 
gravity, do propagate in higher dimensions (known as bulk). Hence the presence of the \KR field brings in modifications to the effective gravitational field equations on any four dimensional hypersurface (referred to as the brane) embedded in the five dimensional bulk \cite{Shiromizu:1999wj,Harko:2004ui,Dadhich:2000am,Maartens:2001jx,Chakraborty:2014xla,Chakraborty:2014fva,Chakraborty:2015bja}. We will study the modified Buchdahl's limit in this scenario as well. The main purpose of this work is precisely to explore these modifications and hence to understand the departure from \gr\ that \KR field as well as existence of higher dimensions can bring in the stellar structures.

The paper can broadly be divided into three sections. In \ref{Buch_Tor_Sec_02} we will discuss the basic mathematical framework of \KR field coupled to gravity and possible effects due to extra dimensions. The formalism developed in \ref{Buch_Tor_Sec_02} will be applied to the derivation of Buchdahl's limit in the context of \KR field in four spacetime dimensions in \ref{Buch_Tor_Sec_03}, while \KR field in higher spacetime dimensions will be discussed in \ref{Buch_Tor_Sec_04}. Finally we conclude with a discussion on our results. 

We will set the fundamental constants $c=1=\hbar$ and will work within the convention of mostly positive signature. All uppercase Roman letters stand for higher dimensional spacetime indices, while lowercase Roman letters indicate spatial indices. On the other hand, Greek letters are used to label the four dimensional spacetime indices. 
\section{Gravity with \KR field: Basic formalism}\label{Buch_Tor_Sec_02}

In this section we will briefly elaborate on the gravitational dynamics in presence of the \KR field. As mentioned earlier, the \KR field is an antisymmetric second rank tensor $B_{AB}$, while the field strength of the \KR field is being denoted by $H_{PQR}=\partial _{[P}B_{Q R]}$\footnote{Note that $\nabla _{[P}B_{QR]}=(1/3)\left\{\nabla _{P}B_{QR}+\nabla _{R}B_{PQ}+\nabla _{Q}B_{RP}\right\}=\partial _{[P}B_{Q R]}$, since all the affine connections cancel out each other due to antisymmetry of the \KR field $B_{PQ}$.}. Alike electrodynamics the action for the \KR field is taken to be square of the field strength. Thus the complete action for \KR field with gravity in $D$ spacetime dimensions takes the following form,
\begin{align}\label{Buch_T_Eq_01}
\mathcal{A}=\int d^{D}x\sqrt{-g}\left[\frac{R}{16\pi G_{D}}-\frac{1}{12}H_{ABC}H^{ABC}+L_{\rm matter}\right]~,
\end{align}
where $G_{D}$ is the D dimensional gravitational constant and $L_{\rm matter}$ consists of any additional matter fields that may be present. The factor $(-1/12)$ ensures that in our signature convention the kinetic term in local inertial frame appears as $(1/2)(\partial _{t}B_{ij})^{2}$. Even though the \KR field in $D$ spacetime dimensions has a total of $D(D-1)/2$ independent components, the actual number of propagating degrees of freedom are lesser in number. This arises from the fact that time derivatives of only the spatial components of the \KR field appear in the Lagrangian. Thus among the total $D(D-1)/2$ independent components only the $(D-1)(D-2)/2$ will represent the propagating degrees of freedom. However one has to take into account of the additional gauge symmetry present in the system, i.e., there is a transformation $B_{PQ}\rightarrow B_{PQ}+\nabla_{P}\xi _{Q}-\nabla _{Q}\xi _{P}$, which keeps the Lagrangian invariant. The gauge field for the spatial part seemingly has $(D-1)$ degrees of 
freedom and hence the total number of degrees of freedom would become $(D-1)(D-4)/2$. There still exists one more scalar degree of freedom, which appears if we change the gauge field by $\xi _{P}\rightarrow \xi _{P}+\partial _{P}\phi$. Thus actual number of degrees of freedom would be $\{(D-1)(D-4)/2\}+1$. Thus in four dimensions the \KR field has a single degree of freedom, while for $D=5$, the number of degrees of freedom becomes three \cite{SenGupta:2001cs,Majumdar:1999jd}.

Having discussed some of the basic properties of the \KR field, let us now inquire how it may affect the dynamics of gravity. For that purpose the most important ingredient is the gravitational field equations, which can be obtained by varying the action in \ref{Buch_T_Eq_01} with respect to the metric, leading to,
\begin{align}
G_{AB}&=8\pi G_{D}\left\{T^{\rm (KR)}_{AB}+T_{AB}^{\rm (matter)}\right\};
\label{Buch_T_Eq_02a}
\\
T^{\rm (KR)}_{AB}&=\frac{1}{6}\left[3H_{APQ}H_{B}^{~PQ}-\frac{1}{2}\left\{H_{PQR}H^{PQR}\right\}g_{AB} \right];\qquad
T^{\rm (matter)}_{AB}=-\frac{2}{\sqrt{-g}}\frac{\delta \left(\sqrt{-g}L_{\rm matter}\right)}{\delta g^{AB}}~.
\label{Buch_T_Eq_02b}
\end{align}
Here $T_{AB}^{\rm (KR)}$ and $T_{AB}^{\rm (matter)}$ respectively corresponds to energy momentum tensor for the \KR field and any additional matter field that may be present in the system. One can also obtain the corresponding field equations for the \KR field by varying the action with respect to $B_{PQ}$ leading to $\nabla _{A}H^{ABC}=0$. Given the field equation for the \KR field it is instructive to prove conservation of the respective energy momentum tensor, which can be derived along the following lines,
\begin{align}
\nabla _{A}T^{A~(\textrm{KR})}_{B}&=\frac{1}{6}\Big[3\left\{\nabla _{A}H^{APQ}\right\}H_{BPQ}
+3H^{APQ}\left\{\nabla _{A}H_{BPQ}\right\}-H^{PQR}\nabla _{B}H_{PQR}\Big]
\nonumber
\\
&=\frac{1}{6}\Big[H^{APQ}\left\{\nabla _{A}H_{BPQ}-\nabla _{Q}H_{ABP}+\nabla_{P}H_{QAB}\right\}-H^{PQR}\nabla _{B}H_{PQR}\Big]
=0~.
\end{align}
Here in the first line we have used the field equation for the \KR field, while in the second line we have used the following result satisfied by the \KR field strength, namely, $\nabla _{[A}H_{PQR]}=0$. The proof of the previous statement is analogous to that of electromagnetism and follows from the complete antisymmetry of the \KR field strength. 

In what follows we will mainly be interested in four dimensional spacetimes with spherical symmetry. This can be achieved along two possible avenues --- (a) One can start from a four dimensional action, which can be obtained by setting $D=4$ in \ref{Buch_T_Eq_01} or (b) Starting from a five dimensional spacetime but then projecting the gravitational field equations on a four dimensional hypersurface. In this work we will explore both these situations, following mainly \cite{Majumdar:1999jd,SenGupta:2001cs,Shiromizu:1999wj,Chakraborty:2014xla,
Chakraborty:2014fva,Mukhopadhyaya:2002jn}, to achieve our goal to derive the possible modifications in the Buchdahl's limit.

We start with the first possibility, i.e., the \KR field in four spacetime dimensions which is described by static and spherically symmetric metric ansatz. The line element fit for our purpose, expressing static and spherical symmetry becomes,
\begin{align}\label{Buch_T_Eq_03}
ds^{2}=-e^{\nu(r)}dt^{2}+e^{\lambda(r)}dr^{2}+r^{2}\left(d\theta ^{2}+\sin ^{2}\theta d\phi ^{2}\right)~,
\end{align}
where $\nu(r)$ and $\lambda(r)$ are arbitrary functions of the radial coordinate that we need to determine through the gravitational field equations. Returning back to the \KR field as we have already emphasized, there is a single independent degree of freedom. This enables one to write down the \KR field tensor $H_{\mu \nu \rho}$ in terms of a scalar field such that, $H^{\mu \nu \rho}=\epsilon ^{\mu \nu \rho \sigma}\partial _{\sigma}\Phi$ \footnote{Note that, since $\epsilon ^{\mu \nu \rho \sigma}=(-1/\sqrt{-g})[\mu \nu \rho \sigma]$, where $[\mu \nu \rho \sigma]$ is the completely antisymmetric object, it immediately follows that, $\nabla _{\mu}H^{\mu \nu \rho}=(-1/\sqrt{-g})[\mu \nu \rho \sigma]\nabla _{\mu}\nabla _{\sigma}\Phi=0$.}. 

In the context of spherical symmetry, the above scalar degree of freedom (also known as axion) becomes a function of radial coordinate alone, i.e., $\Phi=\Phi(r)$. Thus only the $H^{023}$ element will contribute, leading to: $H^{023}=\epsilon ^{0231}\Phi'(r)$, where `prime' denotes derivative with respect to the radial coordinate. Since $\epsilon ^{\mu \nu \rho\sigma}$ involves a $1/\sqrt{-g}$ factor, it follows that $H^{023}$ can be written as $f(r)/\sin \theta$, with $f(r)$ being an arbitrary function of the radial coordinate. This can also be verified by solving the field equation for $H^{\mu \nu \rho}$ directly. With $H^{023}$ as the only non-zero element, the energy momentum tensor for the \KR field turns out to have the following components,
\begin{align}
T^{0~(\textrm{KR})}_{0}&=\frac{1}{6}\Big[6H^{023}H_{023}-\frac{1}{2}\times \left\{6H^{023}H_{023}\right\}\Big]=\frac{1}{2} H^{023}H_{023}=-e^{-\nu(r)}\frac{H_{023}^{2}}{2r^{4}\sin ^{2}\theta}\equiv-h(r)^{2}~,
\label{Buch_T_Eq_04a}
\\
T^{1~(\textrm{KR})}_{1}&=-\frac{1}{2}H^{023}H_{023}=h(r)^{2}=-T^{2~(\textrm{KR})}_{2}=-T^{3~(\textrm{KR})}_{3}~.
\label{Buch_T_Eq_04b}
\end{align}
This suggests that the energy momentum tensor arising due to the \KR field in a spherically symmetric context can actually be expressed as that of a perfect fluid, with the following structure, $\textrm{diag}\left\{-\rho_{\rm KR}(r),p_{\rm KR}(r),-p_{\rm KR}(r),-p_{\rm KR}(r)\right\}$, where $\rho _{\rm KR}(r)=h(r)^{2}=p_{\rm KR}(r)$. It is clear that the energy density is positive definite, since the \KR field is necessarily real. Surprisingly, the transverse (or, angular) part of the energy momentum tensor depicts the existence of a negative pressure. Due to this fact even though the \KR field satisfies the weak energy condition, i.e., $T_{ab}u^{a}u^{b}>0$, it also satisfies, $\{T_{ab}-(1/2)Tg_{ab}\}u^{a}u^{b}=0$. Thus \KR field in spherically symmetric context in four spacetime dimension marginally satisfies the strong energy condition. Incidentally for \KR field, the field equation $\nabla _{\mu}H^{\mu \rho \sigma}=0$ merely says that it can be written in terms of an axionic field, but the identity $\
nabla _{[\mu}H_{\alpha \beta \rho]}=0$ (or, equivalently $\nabla _{\mu}T^{\mu \nu}=0$) provides the differential equation satisfied by $h(r)$. 

Let us now briefly mention about the existing results in the literature in the context of scalar-tensor theory to provide a comparison with the results appearing in our approach. An initial attempt to understand the Buchdahl's limit in the context of Brans-Dicke theory (a particular scalar tensor model) was presented in \cite{Tsuchida:1998jw}, where a conformal transformation was used to transform the Brans-Dicke action in Jordan frame (i.e., when the action has coupling between Ricci scalar $R$ and the scalar field $\phi$) to that in Einstein frame (where the Ricci scalar has no coupling with the scalar field). Then it was demonstrated that the Buchdahl's limit in Brans-Dicke theory is larger compared to the general relativistic scenario. Later on several results have confirmed the above claim in various different scalar tensor models 
\cite{Burikham:2016cwz,Tsuneishi:2005um,Horbatsch:2010hj,Pani:2011xm,Horbatsch:2012hla,
Tsuchida:1998jw}. Thus in this work we provide yet another origin to arrive at a scalar tensor description of gravity, namely using Kalb-Ramond field. In the later part of this work we will explore whether this model also shares the same scenario as far as the Buchdahl's limit is concerned. Since a direct correspondence with earlier results will certainly bolster our claims, we will explore this connection with earlier works in \ref{Buch_Tor_Sec_03} in a detailed manner.

In the present context, given the contribution from the \KR field as expressed in \ref{Buch_T_Eq_04a} and \ref{Buch_T_Eq_04b} respectively, one can write down the corresponding gravitational field equations in the context of spherical symmetry. Further as the stellar interior is concerned, the normal matter is taken to be perfect fluid with energy-momentum tensor $\textrm{diag}\{-\rho(r),p(r),p(r),p(r) \}$. This completes our preliminary discussion and yields the necessary ingredients that we will require in later sections while discussing the effect of \KR field on stellar structure in four spacetime dimensions.  

Before concluding this section, let us address the corresponding situation in the brane world scenario, where both gravity and the \KR field live in five dimensional bulk, while we are interested in the gravitational dynamics on the four dimensional brane. There are several ways of handling this issue. For example, one may wish to average the bulk Einstein's equations over the extra dimension and hence arrive at a gravitational equation on the brane (this method was adopted in \cite{Csaki:1999mp}), otherwise one may wish to project the five dimensional equations on a four dimensional hypersurface. Except these two there are other perturbative schemes available to determine the metric on the brane, inheriting bulk corrections \cite{Kanno:2002iaa,Shiromizu:2002qr}. Nonetheless we will follow the second pathway of projecting the bulk equations in this work, which was first developed in \cite{Shiromizu:1999wj}. Later on, there have been numerous works based on this projective scheme, for a small representative 
set, see \cite{Shiromizu:1999wj,Harko:2004ui,Dadhich:2000am,Maartens:2001jx,Chakraborty:2014xla,
Chakraborty:2014fva,Chakraborty:2015bja,Chakraborty:2015taq}. Since details of this procedure are well established and discussed at good lengths in the above works we will concentrate here by illustrating the basic ingredients which will be necessary for our later purposes. 

The bulk field equations presented in \ref{Buch_T_Eq_02a} can be appropriately projected on the brane hypersurface by using the projector $h^{A}_{B}=\delta ^{A}_{B}-n^{A}n_{B}$, where $n_{A}$ is normal to the brane hypersurface. Using the projector $h^{A}_{B}$ and incorporating the Gauss-Codazzi relation it is possible to write down the brane curvature tensor in terms of the bulk one. In this process one also derives the brane Ricci tensor and hence the brane Ricci scalar in terms of the bulk curvature components. Use of all these results relating brane curvature tensors to bulk curvature tensors enable one to write down the bulk gravitational field equations in terms of curvatures on the four dimensional brane hypersurface with additional contributions inherited from the bulk. In particular, from a purely gravitational point of view the bulk Einstein tensor $G_{AB}$ will map into $~^{(4)}G_{\mu \nu}+E_{\mu \nu}$, where $~^{(4)}G_{\mu \nu}$ is the standard Einstein tensor on the brane and $E_{\mu \nu}=W_{\mu 
A \nu B}n^{A}n^{B}$ is the additional contribution inherited from the bulk, dependent on the bulk Weyl tensor $W_{\mu A \nu B}$ \cite{Shiromizu:1999wj,Harko:2004ui,Dadhich:2000am,Maartens:2001jx}. Similarly, the \KR field present in the bulk (affecting the bulk Einstein's equation through the bulk energy momentum tensor presented in \ref{Buch_T_Eq_02b}) also gets projected on the brane hypersurface and a particular combination of this projection will act as the source of four-dimensional effective gravitational field equations along with normal matter, which of course is confined to four dimensions \cite{gravitation}. In this context, the effect of extra dimensions as well as that of \KR field modifies the four dimensional gravitational field equations as,
\begin{align}
~^{(4)}G_{\mu \nu}+E_{\mu \nu}=8\pi G_{4}\left\{T_{\mu \nu}^{\rm (matter)}+\Pi_{\mu \nu}^{\rm (matter)}+~^{(4)}T_{\mu \nu}^{\rm (KR)}\right\}~,
\end{align}
where the physical interpretation of the terms appearing on the left hand side has been discussed earlier, in particular $E_{\mu \nu}$ is the projection of the bulk Weyl tensor on the brane hypersurface. On the other hand, the right hand side has contribution from three parts --- (a) four dimensional matter field characterized by $T_{\mu \nu}^{\rm (matter)}$, (b) energy momentum tensor obtained by projecting the \KR field on the brane, denoted as $~^{(4)}T_{\mu \nu}^{\rm (KR)}$ and (c) energy-momentum tensor $\Pi _{\mu \nu}^{\rm (matter)}$, which is quadratic in $T_{\mu \nu}^{\rm (matter)}$. The explicit expression for the brane energy momentum tensor $~^{(4)}T_{\mu \nu}^{\rm (KR)}$ originated from the bulk \KR field becomes,
\begin{align}
~^{(4)}T_{\mu \nu}^{\rm KR}&=\frac{2}{3}\frac{G_{5}}{G_{4}}\left[T_{\mu \nu}+\left(T_{AB}n^{A}n^{B}-\frac{1}{4}T_{A}^{A}\right)g_{\mu \nu} \right]
\nonumber
\\
&=\frac{2}{3}\frac{G_{5}}{G_{4}}\Bigg[\frac{1}{2}H_{\mu \alpha \beta}H_{\nu}^{~\alpha \beta}
-\frac{3}{16}\left\{H_{\alpha \beta \rho}H^{\alpha \beta \rho}\right\}g_{\mu \nu}\Bigg]~.
\end{align}
The first equality follows from the projection scheme and has been elaborated in \cite{gravitation}. Moreover $T_{AB}$ appearing on the right hand side of the first expression is the bulk energy momentum tensor for the \KR field, expressed as in \ref{Buch_T_Eq_02b}. In order to arrive at the second expression we have assumed that the \KR field $B_{AB}$ is independent of the extra dimensional coordinate as well as $n^{A}B_{AB}=0$. Both of which are natural from the perspective of a brane observer and can be motivated along the following lines. First of all there are large number of gauge freedoms present in the \KR field, namely $B_{\mu \nu}\rightarrow B_{\mu \nu}+(\partial _{\mu}A_{\nu}-\partial _{\nu}A_{\mu})$, where $A_{\mu}$ is an arbitrary vector field. Using this gauge freedom one can set $n^{A}B_{AB}=0$, known in the literature as the Randall-Sundrum gauge (since it was first used in the context of gravitational perturbation in \cite{Randall:1999vf}). The other condition (i.e., $B_{\mu \nu}$ is 
independent of extra dimension) can be argued from the \KK mode decomposition of the \KR field, where it turns out that the massless \KK mode have no dependence on extra dimension and is purely a function of the brane coordinates \cite{Mukhopadhyaya:2002jn}. Since we are interested in the lowest lying, i.e., massless mode alone, the \KR field cannot have any extra dimension dependence, thereby justifying the previous assumptions.

We would also like to point out that since the above energy momentum tensor is not derivable from a four dimensional action, the conservation of $~^{(4)}T_{\mu \nu}^{\rm (KR)}$ is not immediate from the field equation for the \KR field. Thus one generally treats the field equation for the \KR field separately and considers the total combination $-E_{\mu \nu}+~^{(4)}T_{\mu \nu}^{\rm (KR)}+\Pi _{\mu \nu}^{\rm (matter)}$ to be conserved.

The above energy momentum tensor $~^{(4)}T_{\mu \nu}^{\rm KR}$ surprisingly has some very interesting properties, e.g., if we consider the static and spherically symmetric situation then only $H_{023}$ (or, $H^{023}$) component of the full \KR field strength will be non-zero. In this case the components of this induced brane energy momentum tensor becomes,
\begin{align}
~^{(4)}T_{0}^{0~(\textrm{KR})}&=\frac{2}{3}\frac{G_{5}}{G_{4}}\Bigg[\frac{1}{2}\left\{2H_{023}H^{023}\right\}
-\frac{3}{16}\left\{6H_{023}H^{023}\right\}\Bigg]=-\frac{1}{12}\frac{G_{5}}{G_{4}}H_{023}H^{023}
\nonumber
\\
&=\frac{e^{-\nu(r)}}{12r^{4}\sin ^{2}\theta} \sqrt{\frac{6}{8\pi G_{4}\lambda_{\rm T}}} H_{023}^{2}
\equiv \frac{\tilde{h}(r)^{2}}{\sqrt{8\pi G_{4}\lambda_{\rm T}}}=~^{(4)}T_{2}^{2~(\textrm{KR})}=~^{(4)}T_{3}^{3~(\textrm{KR})};
\\
~^{(4)}T_{1}^{1~(\textrm{KR})}&=\frac{2}{3}\frac{G_{5}}{G_{4}}\Bigg[
-\frac{3}{16}\left\{6H_{023}H^{023}\right\}\Bigg]=\frac{3e^{-\nu(r)}}{4r^{4}\sin ^{2}\theta}\frac{G_{5}}{G_{4}}H_{023}^{2}
=\frac{9\tilde{h}(r)^{2}}{\sqrt{8\pi G_{4}\lambda _{\rm T}}}~,
\end{align}
where $\lambda_{\rm T}=6(G_{4}/8\pi G_{5}^{2})$ is the brane tension. Thus we immediately observe that the induced energy momentum tensor on the brane has \emph{negative} energy density. One can immediately check that, $T_{ab}u^{a}u^{b}\propto -\tilde{h}^{2}$, for static observers, while $\{T_{ab}-(1/2)Tg_{ab}\}u^{a}u^{b}\propto -\tilde{h}^{2}+(1/2)(12\tilde{h}^{2})=5\tilde{h}^{2}$. Thus the induced energy momentum tensor on the brane from a bulk \KR field \emph{violates} the weak energy condition but does satisfy the strong energy condition. This should not come as a surprise, as there have been numerous instances in the literature, in various other contexts where weak energy conditions are being violated on the brane while they are being satisfied in the bulk (which is true for the scenario presented here as well). For example, in the context of a black hole on the brane it was argued that the induced energy density in the brane from the bulk must be negative to ensure attractive nature of gravity on the 
brane \cite{Dadhich:2000ba}. Furthermore, the existence of negative energy density on the brane hypersurface has appeared in numerous other contexts, e.g., (a) Kaluza-Klein reduction to a lower dimensional hypersurface \cite{Wesson:1992bn}, (b) trajectory of a test particle in a lower dimensional hypersurface and appearance of an extra force \cite{Youm:2000ax}, (c) due to topological defects on the hypersurface possibly created by a moving black hole \cite{Frolov:2003mc,Barcelo:2000ta} or may be due to some specific compactification scheme \cite{Hertog:2003ru} (see also  \cite{Vollick:2000uf}). Thus the above provides one more instance to violate weak energy conditions on the brane, namely through the bulk \KR field. Hence by no means this is unusual, it merely provides a new pathway for understanding the energy conditions in the context of braneworld scenario. Having discussed the physics involved as well as basic mathematical formalism, we will next consider the stellar structure and hence the Buchdahl's 
limit in these scenarios. 
\section{\KR field in four dimensions and limit on stellar structure}\label{Buch_Tor_Sec_03}

The basic field equations describing both gravity and \KR field in four spacetime dimensions have been elaborated in the previous section. In particular we have explicitly demonstrated that in this context there exist a single degree of freedom in the \KR field that we should worry about. In the case of static spacetime with spherical symmetry the field equations for gravity as well as that of \KR field simplify considerably. The gravity sector is being determined by two unknowns $\lambda(r)$ and $\nu(r)$ appearing in the spacetime geometry through \ref{Buch_T_Eq_03}, while information regarding \KR field is essentially contained in the unknown function $h(r)$, introduced in \ref{Buch_T_Eq_04a}. Considering the interior of a stellar object to be filled with perfect fluid with energy density $\rho(r)$ and pressure $p(r)$, the gravitational field equations become,
\begin{align}
e^{-\lambda}\left(\frac{1}{r^{2}}-\frac{\lambda '}{r}\right)-\frac{1}{r^{2}}&=8\pi G_{4} \left(-\rho-h^{2}\right)~,
\label{Eq_lambda_01}
\\
e^{-\lambda}\left(\frac{\nu'}{r}+\frac{1}{r^{2}}\right)-\frac{1}{r^{2}}&=8\pi G_{4} \left(p+h^{2}\right)~,
\label{Eq_nu_01}
\\
\frac{1}{2}e^{-\lambda}\left(\nu''+\frac{\nu'^{2}}{2}+\frac{\nu'-\lambda'}{r}-\frac{\nu'\lambda'}{2} \right)&=8\pi G_{4} \left(p-h^{2}\right)~.
\label{Eq_Extra_01}
\end{align}
On the other hand, the conservation equation for the fluid as well as the field equation for the \KR field takes the following simple form in the context of static and spherically symmetric spacetime,
\begin{align}
p'+\frac{\nu'}{2}\left(p+\rho\right)&=0~,
\label{Eq_p_01}
\\
h'+\frac{\nu'}{2}h+\frac{2}{r}h&=0~.
\label{Eq_h_01}
\end{align}
One can solve for these equations if the energy density $\rho(r)=\rho_{\rm c}$, is a constant. Then \ref{Eq_p_01} can be integrated to yield, $\exp(-\nu/2)=A(\rho_{\rm c}+p)$. Given this one can also integrate \ref{Eq_h_01} to obtain the contribution from the \KR field as: $h(r)=(1/r^{2})\exp(-\nu/2)=(A/r^{2})(\rho_{\rm c}+p)$. From this expression it is evident that $h(r)$ decreases with an increase of the radial coordinate $r$. Hence the effective density $\rho+h^{2}$ also decreases as the surface of the star is being approached. We will use this result later on. However the above exact solutions in the case of constant density does not help much, since $p(r)$ is still undetermined, which essentially makes $h(r)$ an arbitrary function of the radial coordinate. 

Interestingly, the three Einstein's equations presented above are not independent, given any two of them along with \ref{Eq_p_01} and \ref{Eq_h_01} one can arrive at the remaining one. We will demonstrate this feature in an explicit manner, since it will provide an important relation which will be useful in our derivation of Buchdahl's limit. We will take \ref{Eq_lambda_01} and \ref{Eq_nu_01} as the two independent equations and shall derive \ref{Eq_Extra_01} from them, where \ref{Eq_p_01} and \ref{Eq_h_01} will be used extensively. The demonstration goes as follows, one starts by differentiating \ref{Eq_nu_01}, which leads to,
\begin{align}\label{Buch_T_Eq_05}
e^{-\lambda}\left(\frac{\nu''}{r}-\frac{\nu'}{r^{2}}-\frac{2}{r^{3}}-\frac{\lambda'\nu'}{r}-\frac{\lambda '}{r^{2}}\right)+\frac{2}{r^{3}}=8\pi G_{4} \left(p'+2hh'\right)~.
\end{align}
Using the conservation equation for the fluid and the \KR field from \ref{Eq_p_01} and \ref{Eq_h_01}, one can evaluate the right hand side of \ref{Buch_T_Eq_05}, leading to,
\begin{align}
8\pi G_{4} \left(p'+2hh'\right)&=8\pi G_{4} \left[-\frac{\nu'}{2}\left(p+\rho\right)+2h\left(-\frac{\nu'}{2}h-\frac{2}{r}h\right) \right]
\nonumber
\\
&=e^{-\lambda}\left(-\frac{\nu'\lambda'}{2r}-\frac{\nu'^{2}}{2r} \right)-\frac{32\pi G_{4}}{r} h^{2}~.
\end{align}
Substitution of this particular expression back to \ref{Buch_T_Eq_05} leads to,
\begin{align}
-32\pi G_{4} h^{2}=e^{-\lambda}\left(\nu''+\frac{\nu'^{2}}{2}-\frac{\lambda'\nu'}{2}+\frac{\nu'-\lambda'}{r}\right)
-16\pi G_{4}\left(p+h^{2}\right)~,
\label{Eq_Extra_02}
\end{align}
where in the last line we have used one of the Einstein's equations, namely \ref{Eq_nu_01}. It is evident that the above equation when divided by a factor of two coincides with \ref{Eq_Extra_01} and hence the third Einstein's equation is redundant. Even then there is one ambiguity present in the system, which is worth mentioning. There are four independent differential equations governing the behaviour of this particular system, while there are five unknowns: $\nu(r),\lambda(r),h(r),p(r)$ and $\rho(r)$. This problem is generally circumvented by assuming an equation of state for the perfect fluid, which we will not need here. 

We will now proceed further and shall determine the fundamental limit to the stellar structure, known in the literature as the Buchdahl's limit, following its discovery by Buchdahl in the context of general relativity. In the remaining part of this section we will derive the Buchdahl's limit in presence of the \KR field explicitly. It is clear that one can integrate out \ref{Eq_lambda_01}, resulting into,
\begin{align}\label{Buch_T_Eq_06}
e^{-\lambda}=1-\frac{2G_{4}m(r)}{r};\qquad m(r)=\int _{0}^{r}dr~4\pi r^{2}\left(\rho+h^{2}\right)~.
\end{align}
Since $h^{2}>0$, it is clear that the total gravitational mass experienced outside the star is larger than the actual matter density present inside, with the extra gravitating mass coming from the KR field strength. Let us now derive the Buchdahl's limit and for that let us start from \ref{Eq_Extra_02} and rewrite the same as,
\begin{align}\label{Eq_Extra_03}
2r\nu''+r\nu'^{2}-r\lambda '\nu'-2\nu'=\frac{4}{r}\left(1-e^{\lambda}\right)+2\lambda'-64\pi G_{4} rh^{2}e^{\lambda}~.
\end{align}
One can further use the following two identities
\begin{align}
\frac{d}{dr}\left[\frac{1}{r}e^{-\lambda/2}\frac{de^{\nu/2}}{dr}\right]
&=\frac{e^{(\nu-\lambda)/2}}{4r^{2}}\left[2r\nu''+r\nu'^{2}-2\nu'-r\nu'\lambda'\right]~,
\label{Eq_Torso_New03a}
\\
\frac{d}{dr}\left[\frac{1-e^{-\lambda}}{2r^{2}}\right]&=\frac{e^{-\lambda}}{2r^{3}}\left[r\lambda'-2\left(e^{\lambda}-1\right)\right]~,
\label{Eq_Torso_New03b}
\end{align}
to rewrite \ref{Eq_Extra_03} as,
\begin{align}\label{Buch_T_Eq_07}
e^{-(\nu+\lambda)/2}\frac{d}{dr}\left[\frac{1}{r}e^{-\lambda/2}\frac{d}{dr}e^{\nu/2} \right]=\frac{d}{dr}\left(\frac{1-e^{-\lambda}}{2r^{2}}\right)-\frac{16\pi G_{4}}{r}h^{2}~.
\end{align}
At this stage one puts forward some sensible requirements, e.g., the average density $\rho_{\rm av}=m(r)/r^{3}$ should decrease outwards. Even though the average density involves contribution from the \KR field, since the \KR field strength itself decreases outwards the above condition will be trivially satisfied. Further, given the form of $e^{-\lambda}$ as in \ref{Buch_T_Eq_06}, it is clear that the first term on the right hand side of \ref{Buch_T_Eq_07} is essentially $d\rho_{\rm av}/dr$. Since we have already assumed that average density decreases outwards, it is clear that the right hand side of \ref{Buch_T_Eq_07} is negative, leading to,
\begin{align}
\frac{d}{dr}\left[\frac{1}{r}e^{-\lambda/2}\frac{d}{dr}e^{\nu/2} \right]\leq 0~.
\end{align}
Integrating the above relation from some radius $r$ within the star to the surface of the star, given by the radius $r_{0}$, we obtain,
\begin{align}
\frac{1}{r}e^{-\lambda/2}\frac{de^{\nu/2}}{dr}\geq \frac{1}{r_{0}}\left[e^{-\lambda/2}
\frac{d}{dr}e^{\nu /2}\right]_{r=r_{0}}=\frac{1}{2r_{0}}e^{(\nu_{0}-\lambda_{0})/2}\nu'_{0}~,
\label{Eq_Torso_New_01}
\end{align}
where quantities with subscript `0' denotes that they are to be evaluated at the surface of the star, located at $r=r_{0}$. At this stage one generally assumes that both the metric and its derivatives are continuous across the surface of the star, i.e., the exterior solution for $\lambda$ and $\nu$ must be equal to the interior solution at the surface along with their derivatives. This prompts to replace the metric and its derivatives at the surface by the respective values associated with the external solution. However, for the moment being we will keep the above structure arbitrary, which will be useful later. Further multiplying both sides of \ref{Eq_Torso_New_01} by $re^{\lambda/2}$ and integrating again from the origin to the surface of the star, we obtain, 
\begin{align}
e^{\nu/2}(r=0)\leq e^{\nu_{0}/2}-\frac{1}{2r_{0}}e^{(\nu_{0}-\lambda_{0})/2}\nu'_{0}\int _{0}^{r_{0}}dr~\frac{r}{\sqrt{1-\frac{2G_{4}m(r)}{r}}}~.
\end{align}
As the average density is assumed to be decreasing outwards, it immediately follows that $m(r)/r^{3}>M/r_{0}^{3}$ and thus we will have the above inequality holding more strongly if we replace $m(r)/r$ by $(M/r_{0}^{3})r^{2}$. With this modification, the above inequality becomes,
\begin{align}\label{Eq_Buch_N_01}
e^{\nu/2}(r=0)&\leq e^{\nu_{0}/2}-\frac{1}{2r_{0}}e^{(\nu_{0}-\lambda_{0})/2}\nu'_{0}\int _{0}^{r_{0}}dr~\frac{r}{\sqrt{1-\frac{2G_{4}M}{r_{0}^{3}}r^{2}}}
\nonumber
\\
&=e^{\nu_{0}/2}-\frac{1}{2r_{0}}e^{(\nu_{0}-\lambda_{0})/2}\nu'_{0}\left\{\frac{r_{0}^{2}}{1-e^{-\lambda_{0}}}\right\} \left(1-e^{-\lambda_{0}/2}\right)~.
\end{align}
Since both the pressure and the contribution from the \KR field are positive and finite at the origin it follows that $e^{\nu/2}(r=0)>0$. Applying this result to \ref{Eq_Buch_N_01} we immediately obtain the following inequality,
\begin{align}
e^{\nu_{0}/2}-\frac{1}{2r_{0}}e^{(\nu_{0}-\lambda_{0})/2}\nu'_{0}\left\{\frac{r_{0}^{2}}{1-e^{-\lambda_{0}}}\right\} \left(1-e^{-\lambda_{0}/2}\right)>0~.
\label{Eq_Torso_New_02}
\end{align}
\begin{figure}
\begin{center}
\includegraphics[scale=1]{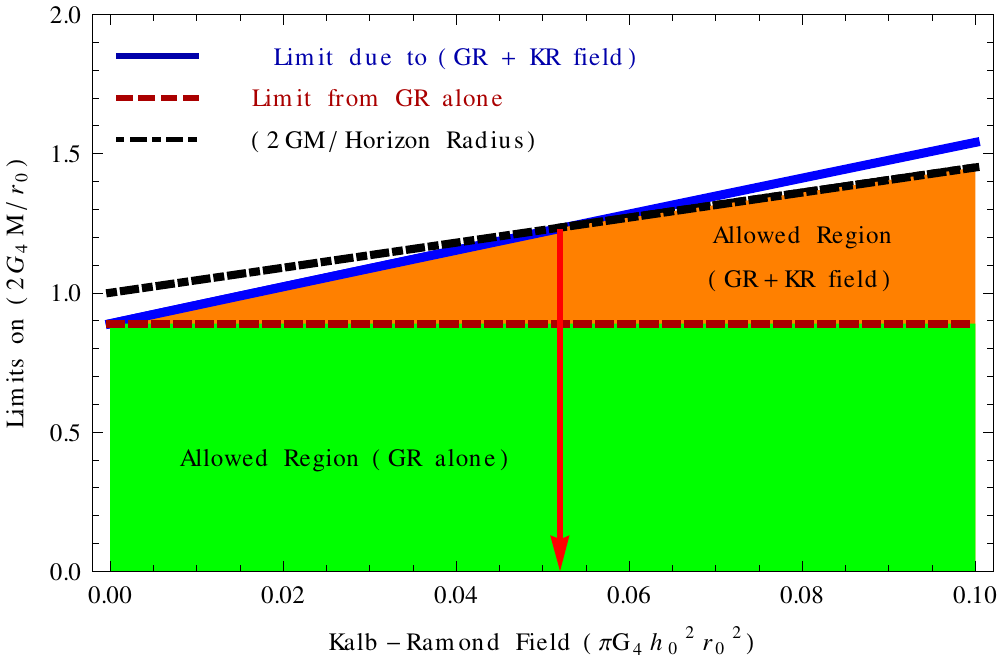}
\end{center}
\caption{Allowed region for $2G_{4}\mathcal{M}/r_{0}$ (region below the blue, thick line) along with the location of the event horizon (black, dot-dashed line) are being depicted as the \KR field makes its appearance in the picture. It is clear that the size of the allowed region is initially bounded by the limit on $2GM/r_{0}$ (to the left of the red arrow), while later on it is decided by the location of the horizon (right of the red arrow). The red arrow depicts the point after which the smallest $r_{0}$ becomes smaller than event horizon. As evident the allowed region over and above \gr\ increases as the \KR field strength increases. Thus one can add more mass compared to \gr\ to the stellar structure at a given radius if the \KR field is present. This may provide a nice observational window to probe possible existence of the \KR field. See text for more discussions.}
\label{Fig_Buch_01}
\end{figure}
Note that the term $e^{\nu_{0}/2}$ can be removed from the above inequality as $e^{\nu_{0}/2}>0$ as well. Further evaluating \ref{Eq_nu_01} at $r=r_{0}$ we obtain the following expression for $\nu'_{0}$,
\begin{align}
\frac{\nu'_{0}}{2r_{0}}=-\frac{1}{2r_{0}^{2}}+\frac{1}{2}e^{\lambda_{0}}\Big\{\frac{1}{r_{0}^{2}}+8\pi G_{4}h_{0}^{2}\Big\}~.
\end{align}
Substitution of the above expression for $\nu'_{0}/2r_{0}$ in \ref{Eq_Torso_New_02} and subsequent multiplication by $e^{-\lambda _{0}/2}$ leads to the following expression,
\begin{align}
e^{-\lambda_{0}/2}-\frac{r_{0}^{2}}{1-e^{-\lambda_{0}}}\left[\frac{1}{2r_{0}^{2}}\left(1-e^{-\lambda_{0}}\right)
+4\pi G_{4}h_{0}^{2}\right]\left(1-e^{-\lambda_{0}/2}\right)>0~.
\end{align}
Hence one can solve for this inequality for a corresponding bound on $\exp(-\lambda_{0})$, which will certainly be different from the corresponding bound in Schwarzschild spacetime. Simplification of the above inequality leads to a quadratic expression for $(1-e^{-\lambda_{0}})$, whose one root corresponds to negative value of the same and hence can be neglected, while the other root provides the necessary inequality, 
\begin{align}\label{Eq_F_01}
1-e^{-\lambda_{0}}<\frac{4}{9}\left[1-6\pi G_{4}r_{0}^{2} h_{0}^{2}+\sqrt{1+6\pi G_{4}h_{0}^{2}r_{0}^{2}}\right]~.
\end{align}
The corresponding bound on the ADM mass can be obtained by writing down $e^{-\lambda _{0}}$ in terms of the same and using the inequality presented in \ref{Eq_F_01}. In order to obtain $e^{-\lambda _{0}}$ in terms of the ADM mass we need to determine the exterior solution, which to the leading order in the \KR parameter $h_{0}$ becomes
\begin{align}
e^{-\lambda}=1-\frac{2G_{4}\mathcal{M}}{r}+8\pi G_{4}\frac{h_{0}^{2}r_{0}^{4}}{r^{2}}+\mathcal{O}(r^{-3})~,
\label{sol_tor_01}
\\
e^{\nu}=1-\frac{2G_{4}\mathcal{M}}{r}+\mathcal{O}(r^{-3});\qquad h(r)=h_{0}\left(\frac{r_{0}^{2}}{r^{2}}\right)+\mathcal{O}(r^{-3})~,
\label{sol_tor_02}
\end{align}
where $\mathcal{M}$ corresponds to the desired ADM mass associated with the black hole spacetime. Thus evaluating $e^{-\lambda}$ on the surface of the star and its subsequent substitution in \ref{Eq_F_01} provides the desired bound on $\mathcal{M}$ as,
\begin{align}\label{Eq_F_02}
\frac{2G_{4}\mathcal{M}}{r_{0}}<\frac{4}{9}\left[1+12\pi G_{4}r_{0}^{2} h_{0}^{2}+\sqrt{1+6\pi G_{4}h_{0}^{2}r_{0}^{2}}\right]~.
\end{align}
One can easily comprehend the correctness of this result by taking the $h_{0}\rightarrow 0$ limit, for which one immediately recovers the Buchdahl's limit, $G_{4}\mathcal{M}/r_{0}<4/9$. It is clear from the above result that in presence of \KR field the upper bound on $G_{4}/r_{0}$ is larger than $4/9$ (see \ref{Fig_Buch_01}). Thus as the strength of the \KR field increases the compactness limit (or, equivalently the Buchdahl's limit) on the stellar structure also increases compared to that in \gr. However, one point must be emphasized at this stage. Even though $2G_{4}\mathcal{M}/r_{0}$ can become much larger than $8/9$ due to higher and higher values of the \KR field parameter $h_{0}$, it is fundamentally bounded by the black hole horizon. To be precise, if the surface of the star is within the event horizon, then the limit makes no sense. Thus one has to ensure that $r_{0}>r_{\rm h}$, where $r_{\rm h}$ is the location of the horizon obtained from \ref{sol_tor_01}, in order to have a 
sensible stellar structure. Thus for $(2G_{4}\mathcal{M}/r_{0})<(2G_{4}\mathcal{M}/r_{\rm h})$, the bound on stellar structure is provided by $(2G_{4}\mathcal{M}/r_{0})$. This corresponds to the region to the left of the red arrow in \ref{Fig_Buch_01}. While in the opposite scenario, the ultimate bound on stellar structure is provided by the horizon radius as depicted by the region to the right of the red arrow in \ref{Fig_Buch_01}. Finally the choice $\pi G_{4}h_{0}^{2}r_{0}^{2}\sim 0.052$ marks the point where the stellar radius and the horizon radius coincides, as illustrated by the red arrow in \ref{Fig_Buch_01}. As evident, in both these situations one can have stellar configurations having more mass packed into a smaller volume, when compared to the corresponding situation in \gr.

For completeness, we would also like to comment on the connection between the spherically symmetric solution presented above, with those derived earlier in the literature (see, for example \cite{Buchdahl:1959nk,Janis:1970kn,Wyman:1981bd,Husain:1994uj}). In particular, we will be considering the spherically symmetric and static solution obtained in \cite{Buchdahl:1959nk}, in the context of Einstein gravity in presence of a non-trivial scalar field. Some generalization of this solution has been achieved and discussed in \cite{Janis:1970kn,Wyman:1981bd}. While an exact solution in the context of gravitational collapse in presence of a scalar field has been addressed in \cite{Husain:1994uj}. To understand the possible connection of the solution presented in this work with those in the earlier literatures, consider first the solution for the scalar field. The profile of the scalar field was given by $\sim 2\lambda \ln \{1-(4m^{2}/r^{2})\}$ \cite{Buchdahl:1959nk}. Thus to leading order in $r^{-1}$, the 
scalar field solution behaves as, $\sim (8\lambda m^{2}/r^{2})$. As evident from \ref{sol_tor_02}, the leading order behaviour for the scalar field exactly matches with the solution for the \KR field in the present context. Further, an inspection of the $g_{tt}$ component of the metric, when converted to the spherically symmetric form starting from the isotropic coordinates as presented in \cite{Buchdahl:1959nk} reveals that to leading order it scales as, $\sim 1-(2m/r)+\mathcal{O}(1/r^{3})$, which can also be compared with \ref{sol_tor_02}. Thus the static and spherically symmetric solution presented here is indeed consistent with the earlier findings in the literature.

The above bound on ADM mass $\mathcal{M}$ illustrates that given a certain radius $r_{0}$ of a compact stellar object, the maximum mass one can associate with it is \emph{larger} in presence of \KR field. Thus one can pack extra mass into the stellar structure (a similar scenario appears in $f(R)$ gravity as well \cite{Goswami:2015dma}.). This provides an interesting testbed for \KR field. For example, if a compact object (possibly neutron star) is observed whose $\mathcal{M}/r_{0}$ ratio is larger then $4/9$, then it can possibly signal towards the existence of a non-zero \KR field. 

Given the significance of the above result in the astrophysical context, it would be of interest if some comment regarding stability of the solution can be made. For this purpose one need to consider three perturbation modes, namely the scaler modes, vector modes and of course the tensor modes. The scalar perturbation is essentially due to a scalar field, while electromagnetic fields are responsible for the vector perturbations. Finally gravitational perturbations are being represented by the tensor modes. In spherically symmetric background all these perturbations satisfy certain master equation governing the evolution of the perturbations. These equations can generically be written as \cite{Chandrasekhar:1985kt,Nollert:1999ji,Kokkotas:1999bd,Berti:2015itd,
Berti:2009kk,Pani:2013pma} 
\begin{align}\label{pert_tor}
\frac{d^{2}\Psi _{s}}{dr_{*}^{2}}+\left\{\omega ^{2}-V_{s}(r_{*})\right\}\Psi_{s}=0~,
\end{align}
where $\Psi _{s}$ is the perturbation variable and $s=0,1,2$ for scalar, vector and tensor modes respectively. Since the spacetime is static the time dependence has been separated by assuming $e^{\pm i\omega t}$ dependence. The stability of the perturbation mode hinges on the fact that there are no growing modes present, which loosely speaking originates from the positivity of the potential $V_{s}$. For example, in the context of the scalar perturbation the structure of the potential in the present context reads
\begin{align}
V_{0}&=e^{\nu}m^{2}+e^{\nu}\frac{\ell(\ell+1)}{r^{2}}-\frac{1}{2r}\partial _{r}e^{\nu-\lambda}~.
\end{align}
By substituting the expressions for $e^{\nu}$ and $e^{-\lambda}$ from \ref{sol_tor_01} and \ref{sol_tor_02} respectively, we observe that this coincides with the corresponding situation in Schwarzschild spacetime, with higher order corrections from the presence of \KR field. This is because $e^{\nu}$ differs from the Schwarzschild solution only at $\mathcal{O}(1/r^{3})$. A similar consideration will apply to the vector and tensor perturbations as well, where the leading contribution will be from Schwarzschild spacetime with sub-leading corrections due to the \KR field. The \KR field strength $h_{0}$ cannot be a large number (otherwise, solar system tests like bending of light would have observed the same, see \cite{Kar:2002xa}) and hence the stability of the Schwarzschild solution ensures that the corresponding solution discussed in this framework is also stable. However, in order to arrive at a complete picture it is necessary to work through the black hole perturbation theory in its full gory detail, 
which will be addressed elsewhere.

The above results are also in complete agreement with the earlier results in the literature in the context of Buchdahl's limit in scalar tensor theories. In various scalar tensor models of gravity, which mostly are of Brans-Dicke origin, it has been demonstrated that the Buchdahl's limit increases, i.e., one can have more mass at a smaller radius. In particular, in \cite{Tsuchida:1998jw} it was claimed that the Buchdahl's limit in the context of scalar tensor theories can even exceed the value unity, representing black hole horizon in general relativity. From \ref{Fig_Buch_01} it is clear that such a scenario is indeed present in our model as well, with a higher value for the \KR field the Buchdahl's limit can definitely cross the black hole barrier. Furthermore, it was argued in \cite{Tsuchida:1998jw} that if the energy density and pressure associated with the scalar field satisfies the condition $\rho-3p>0$, then the original Buchdahl's limit will be retrieved. One can trivial check that for \KR field such 
a condition can never be satisfied and hence we will always have modifications to the Buchdahl's limit. The above results explicitly demonstrate the robustness of the method presented in this work and are in complete agreement with earlier literatures \cite{Burikham:2016cwz,Tsuneishi:2005um,Horbatsch:2010hj,Pani:2011xm,Horbatsch:2012hla,Tsuchida:1998jw}. This completes our discussion on stellar structure in presence of \KR field in four spacetime dimensions. We will now discuss the corresponding scenario when extra spacetime dimensions are present.
\section{\KR field in induced gravity theory and limits on stellar structure}\label{Buch_Tor_Sec_04}

In the previous section we had discussed in detail how the Buchdahl's limit gets affected in presence of \KR field in four spacetime dimensions. As emphasized earlier, the \KR field being a closed string mode can probe the higher dimensions. Thus it is legitimate to ask, how the above calculation and in particular the Buchdahl's limit gets affected by the presence of \KR field in higher dimensions. We will answer this question by using the effective field equation technique to obtain the gravitational field equations on the brane hypersurface. The presence of extra dimensions will bring an additional parameter, namely the brane tension $\lambda_{\rm T}$ in the picture. In addition there will be several extra pieces in the gravitational field equations on the brane, e.g., effects of the bulk Weyl tensor $E_{\mu \nu}$, projection of the energy momentum tensor of the bulk \KR field etc. Using the results obtained in \ref{Buch_Tor_Sec_02} in the context of spherical symmetry the corresponding field equations in 
the interior of the perfect fluid star become,
\begin{align}
e^{-\lambda}\left(\frac{1}{r^{2}}-\frac{\lambda'}{r}\right)-\frac{1}{r^{2}}&=-8\pi G_{4}\rho-\frac{8\pi G_{4}}{2\lambda_{\rm T}}\rho ^{2}-\frac{6U}{8\pi G_{4}\lambda _{\rm T}}+\sqrt{\frac{8\pi G_{4}}{\lambda _{\rm T}}}\tilde{h}^{2}~,
\label{Torsion_Star_B1a}
\\
e^{-\lambda}\left(\frac{\nu'}{r}+\frac{1}{r^{2}}\right)-\frac{1}{r^{2}}&=8\pi G_{4}p+\frac{8\pi G_{4}}{2\lambda_{\rm T}}\rho \left(\rho+2p\right)
+\frac{2}{8\pi G_{4}\lambda_{\rm T}}\left(U+2P\right)+9\sqrt{\frac{8\pi G_{4}}{\lambda _{\rm T}}}\tilde{h}^{2}~,
\label{Torsion_Star_B1b}
\\
\frac{1}{2}e^{-\lambda}\left(\nu''+\frac{\nu'^{2}}{2}+\frac{\nu'-\lambda'}{r}-\frac{\nu'\lambda'}{2} \right)&=8\pi G_{4}p +\frac{8\pi G_{4}}{2\lambda _{\rm T}}\rho \left(\rho+2p\right)
+\frac{2}{8\pi G_{4}\lambda _{\rm T}}\left(U-P\right)+\sqrt{\frac{8\pi G_{4}}{\lambda _{\rm T}}}\tilde{h}^{2}~.
\label{Torsion_Star_B1c}
\end{align}
Here $U=-(G_{4}/G_{5})^{2}E_{\mu \nu}u^{\mu}u^{\nu}$ is the ``dark radiation'' term and $P=(1/2)(G_{4}/G_{5})^{2}E_{\mu \nu}(u^{\mu}u^{\nu}-3r^{\mu}r^{\nu})$ is the ``dark pressure'' term originating from the bulk Weyl tensor induced on the brane hypersurface. The vector $u^{\mu}$ corresponds to any timelike vector on the brane hypersurface, while $r_{\mu}$ corresponds to another spacelike vector on the brane hypersurface, such that $u_{\mu}r^{\mu}=0$ \cite{Dadhich:2000am}. From the above set of equations it is clear that in the limit $\lambda _{\rm T}\rightarrow \infty$, one recovers the Einstein's equations with perfect fluid source. Having obtained the Einstein's equations in the context of brane spacetime with Kalb-Ramond field, let us concentrate on the associated conservation relations.

The conservation of perfect fluid energy momentum tensor representing the matter content of the star is again given by the standard expression as in \ref{Eq_p_01}. While the field equation for the \KR field and the conservation of the remaining tensors yield, 
\begin{align}
p'&+\frac{\nu'}{2}\left(p+\rho\right)=0~,
\label{Torsion_Star_B2a}
\\
\tilde{h}'&+\frac{\nu'}{2}\tilde{h}+\frac{2}{r}\tilde{h}=0~,
\label{Torsion_Star_B2b}
\\
\frac{1}{8\pi G_{4}\lambda _{\rm T}}\Bigg\{U'&+2\nu'U+2P'+\nu'P+\frac{6}{r}P\Bigg\}=-\frac{8\pi G_{4}}{2\lambda _{\rm T}}\rho'(\rho+p)+5\sqrt{\frac{8\pi G_{4}}{\lambda _{\rm T}}}\Big[\frac{\nu'}{2}\tilde{h}^{2}+\frac{2}{r}\tilde{h}^{2} \Big]~,
\label{Torsion_Star_B2c}
\end{align}
where \ref{Torsion_Star_B2b} corresponds to $\nabla _{[\mu}H_{\alpha \beta \rho]}=0$, with $\lambda _{T}$ assumed to be finite. Thus \ref{Torsion_Star_B2b} has no \gr\ limit. On the other hand, \ref{Torsion_Star_B2c} originates by using \ref{Torsion_Star_B2a} and \ref{Torsion_Star_B2b} respectively in the conservation of $\{-E_{\mu \nu}+~^{(4)}T_{\mu \nu}^{\rm KR}+\Pi _{\mu \nu}^{\rm matter}\}$. Note that in the limit $\lambda _{\rm T}\rightarrow \infty$, the last conservation relation becomes trivial. 

Given the above set of equations, one can infer some nice properties regarding the system without going into too much details. For example, if the star has constant density $\rho_{\rm c}$, then \ref{Torsion_Star_B2a} and \ref{Torsion_Star_B2b} can be integrated to yield, $\exp(-\nu/2)=A(p+\rho_{\rm c})$, and the \KR field will behave as $h(r)=(1/r^{2})\exp(-\nu/2)=(A/r^{2})(p+\rho_{c})$. Thus in this case as well the contribution from \KR field decreases as one moves towards the surface of the star. Further in this case of constant density star, from \ref{Torsion_Star_B2c} it is clear that the bulk stresses have to be nonzero both inside and outside of the star, since $h(r)$ is non-zero everywhere. This is in sharp contrast to the corresponding situation depicted in \cite{Germani:2001du} and arises solely due to the presence of the \KR field. Moreover, \ref{Torsion_Star_B2c} cannot be integrated directly to provide an expression for $U(r)$ when $P(r)=0$, unlike the situation in \cite{Germani:2001du}, again 
due to the \KR field. Thus we conclude that there will be non-zero Weyl stresses present both inside and outside the stellar object, in presence of non-zero \KR field (see also \cite{Ovalle:2016pwp,Ovalle:2014uwa,Ovalle:2013vna,Casadio:2012rf}). 

It is also possible to argue, using \ref{Torsion_Star_B2a} and \ref{Torsion_Star_B2c}, that the dark pressure $P$ should be positive under reasonable physical assumptions. First of all, as already mentioned earlier, we assume that the energy density $\rho$ and pressure $p$ of the perfect fluid acting as the building material of the star decreases outwards. This suggests that $\rho'$ and $p'$ are both negative. Therefore as evident from \ref{Torsion_Star_B2a} $\nu'>0$, since $(\rho+p)$ is positive definite. We also assume that an identical situation holds for the dark pressure and dark radiation as well, or in other words these two quantities decrease as one reaches the outer region of the stellar structure, implying both $U'$ and $P'$ to be negative \cite{Germani:2001du}. With these reasonable assumptions, let us examine \ref{Torsion_Star_B2c} in some detail. Firstly all the terms on the right hand side of \ref{Torsion_Star_B2c} are positive, thanks to the fact that $\rho'<0$ but $\nu'>0$, while $(\rho+p)$ 
is positive definite. Therefore the left hand side of \ref{Torsion_Star_B2c} should also be positive. However the terms $U'$ and $P'$ are negative, keeping only three terms depending either on $U$ or $P$ linearly to compensate them. In the regime of linear perturbation theory, with (brane curvature/bulk curvature) as the perturbation parameter, one can show that $U=-E_{\mu \nu}u^{\mu}u^{\nu}<0$ \cite{Mukohyama:1999wi,Sasaki:1999mi} and therefore for \ref{Torsion_Star_B2c} to hold it is necessary that $P>0$. A similar conclusion can also be reached by using Big-Bang Nucleosynthesis as well as Cosmic Microwave Background to understand the dark radiation term. Using the current estimates for primordial $^{4}\textrm{He}$ as well as Deuterium to Hydrogen ratio one can safely argue that negative values of $U$ are much more favoured compared to the scenario with $U$ being positive \cite{Ichiki:2002eh}. Thus perturbative estimations as well as observational compatibility support in favour of negative dark radiation. 
Interestingly, the fact that the dark radiation term is negative is important to match the interior solution to the exterior one, since the exterior solutions generically have negative dark radiation \cite{Dadhich:2000am,Germani:2001du}. Hence we can conclude that as far as the current scenario is considered (in particular \ref{Torsion_Star_B2c}), it votes for a positive value of the dark pressure term $P$, a fact which we will use in this work.

Using \ref{Torsion_Star_B1a} one can solve for $e^{-\lambda}$ and hence obtain the mass function in the interior of the stellar structure, leading to,
\begin{align}\label{Eq_N_time}
e^{-\lambda}=1-\frac{2G_{4}m(r)}{r};\qquad m(r)=\int ^{r}_{0}dr~4\pi r^{2}\Bigg\{\rho+\frac{\rho ^{2}}{2\lambda _{\rm T}}+\frac{6U}{(8\pi G_{4})^{2}\lambda _{\rm T}}-\frac{\tilde{h}^{2}}{\sqrt{8\pi G_{4}\lambda _{\rm T}}} \Bigg\}~.
\end{align}
Let us now analyze the structure of the above equation. The first term in the expression for $m(r)$ is the normal matter energy density producing mass, while the second one arises due to the presence of extra dimensions. The third one has a purely geometric origin, namely from the projection of bulk Weyl tensor and the last bit is from the bulk \KR field. The minus sign in front of the \KR field strength ensures that it is effectively lowering the gravitational mass with respect to the situation when the \KR field is absent (i.e., compared to the $\tilde{h}=0$ situation). This is unlike the situation in \ref{Buch_Tor_Sec_03}, where the \KR field appears with a positive sign. Among the other two additional terms in \ref{Eq_N_time}, $\rho^{2}$ is always positive definite, while $U$ can have either positive or negative contribution.

To see that the three gravitational field equations are not independent (they cannot be, as there are only two unknown functions $\lambda$ and $\nu$), we can start with the derivative of \ref{Torsion_Star_B1b} with respect to the radial coordinate resulting into,
\begin{align}\label{F_Buch_Eq_01}
e^{-\lambda}\Big(\frac{\nu''}{r}-\frac{\nu'}{r^{2}}-\frac{2}{r^{3}}-\frac{\lambda'\nu'}{r}-\frac{\lambda'}{r^{2}} \Big)+\frac{2}{r^{3}}
=-\frac{\nu'}{2}\Big[8&\pi G_{4}\left(p+\rho\right)+8\sqrt{\frac{8\pi G_{4}}{\lambda _{\rm T}}}\tilde{h}^{2}+\frac{8\pi G_{4}}{2\lambda _{\rm T}}2\rho\left(p+\rho\right)
\nonumber
\\
&+\frac{2}{8\pi G_{4}\lambda _{\rm T}}\left(4U+2P\right)\Big]-\frac{16}{r}\sqrt{\frac{8\pi G_{4}}{\lambda _{\rm T}}}\tilde{h}^{2}-\frac{12P}{8\pi G_{4}\lambda _{\rm T} r}~.
\end{align}
The term inside square bracket is actually the subtraction of \ref{Torsion_Star_B1a} from \ref{Torsion_Star_B1b} as one can immediately verify. Using this fact along with rearrangement of terms, \ref{F_Buch_Eq_01} finally leads to,
\begin{align}
\frac{1}{r}e^{-\lambda}\left(\nu''+\frac{\nu'^{2}}{2}-\frac{\lambda'\nu'}{2}+\frac{\nu'-\lambda'}{r}\right)
=\frac{2}{r}\left[e^{-\lambda}\left(\frac{\nu'}{r}+\frac{1}{r^{2}}\right)-\frac{1}{r^{2}}\right]-\frac{16}{r}\sqrt{\frac{8\pi G_{4}}{\lambda _{\rm T}}}\tilde{h}^{2}-\frac{12P}{8\pi G_{4}\lambda _{\rm T} r}~.
\label{Torsion_Star_B3}
\end{align}
As \ref{Torsion_Star_B1b} is being substituted for the term within square bracket in the right hand side, one gets back \ref{Torsion_Star_B1c}. Thus \ref{Torsion_Star_B1c} is indeed not an independent equation, but depends on the other two. The main reason behind this derivation is the fact that the above equation is quiet useful for our purpose, namely derivation of Buchdahl's limit. Keeping this in mind one can also rewrite \ref{Torsion_Star_B3} such that the following relation is obtained,
\begin{align}
2r\nu''+r\nu'^{2}-r\lambda '\nu'-2\nu'=\frac{4}{r}\left(1-e^{-\lambda}\right)+2\lambda'
-\frac{24 r P}{8\pi G_{4}\lambda _{\rm T}}e^{\lambda}-32r\sqrt{\frac{8\pi G_{4}}{\lambda _{\rm T}}}\tilde{h}^{2}e^{\lambda}~.
\end{align}
At this stage one can use the two identities introduced in \ref{Eq_Torso_New03a} and \ref{Eq_Torso_New03b} respectively, and then simplifying the resulting expression further, we obtain the following result,
\begin{align}
\exp\left(-\frac{\lambda+\nu}{2}\right)\frac{d}{dr}\left[\frac{1}{r}e^{-\lambda/2}\frac{de^{\nu/2}}{dr}\right]
=\frac{d}{dr}\left(\frac{1-e^{-\lambda}}{2r^{2}} \right)
-\frac{6P}{8\pi G_{4}\lambda _{\rm T} r}-8\sqrt{\frac{8\pi G_{4}}{\lambda _{\rm T}}}\frac{h^{2}}{r}~.
\label{Eq_Torso_N4}
\end{align}
Among the three terms on the right hand side, the last one originating from the bulk \KR field as well the one from dark pressure provides a negative contribution, since following our argument below \ref{Torsion_Star_B2c} it turns out that $P>0$. Another way to justify the positiveness of dark pressure is as follows: note that the configurations for $U$ and $P$ inside the stellar structure can not be arbitrary, as they have to match with the exterior solution as well. The exterior solution requires the pressure to be positive (see, e.g., \cite{Dadhich:2000am,Germani:2001du}), which is mainly due to the origin of this term from bulk Weyl tensor. Thus for reasonable exterior solutions we will expect the ``dark pressure'' to be positive in the interior as well. On the other hand, as far as the first term in the right hand side of \ref{Eq_Torso_N4} is concerned, \ref{Eq_N_time} ensures that it is the rate of change of average density $m(r)/r^{3}$. This requires the dark radiation term to decrease outwards as the 
surface of the star is being approached (which it indeed does, see \cite{Germani:2001du}). Among the other terms in \ref{Eq_N_time}, $\rho$ and $\rho^{2}$ are both decreasing functions of the radial coordinate and as already pointed out, the \KR field strength $\tilde{h}(r)$ decreases as one moves radially outwards. Thus one can safely impose the assumption that average density should decrease outward. This ensures that the following inequality is being satisfied,
\begin{align}
\frac{d}{dr}\left[\frac{1}{r}e^{-\lambda/2}\frac{de^{\nu/2}}{dr}\right]<0~.
\end{align}
Given this, one can proceed as in the previous section and following identical steps one finally arrives at the desired inequality involving $e^{\lambda}$ alone, such that,
\begin{align}
e^{-\lambda_{0}/2}-\left[\frac{1}{2r_{0}^{2}}\left(1-e^{-\lambda_{0}}\right)+\frac{8\pi G_{4}}{2\lambda _{\rm T}}\frac{\rho_{0}^{2}}{2}
+\frac{2}{8\pi G_{4}\lambda _{\rm T}}\left(\frac{U_{0}}{2}+P_{0}\right)+\frac{9}{2}\sqrt{\frac{8\pi G_{4}}{\lambda _{\rm T}}}\tilde{h}_{0}^{2}\right]\int _{0}^{r_{0}}dr~re^{\lambda/2}
>0~.
\end{align}
\begin{figure}
\begin{center}
\includegraphics[scale=0.75]{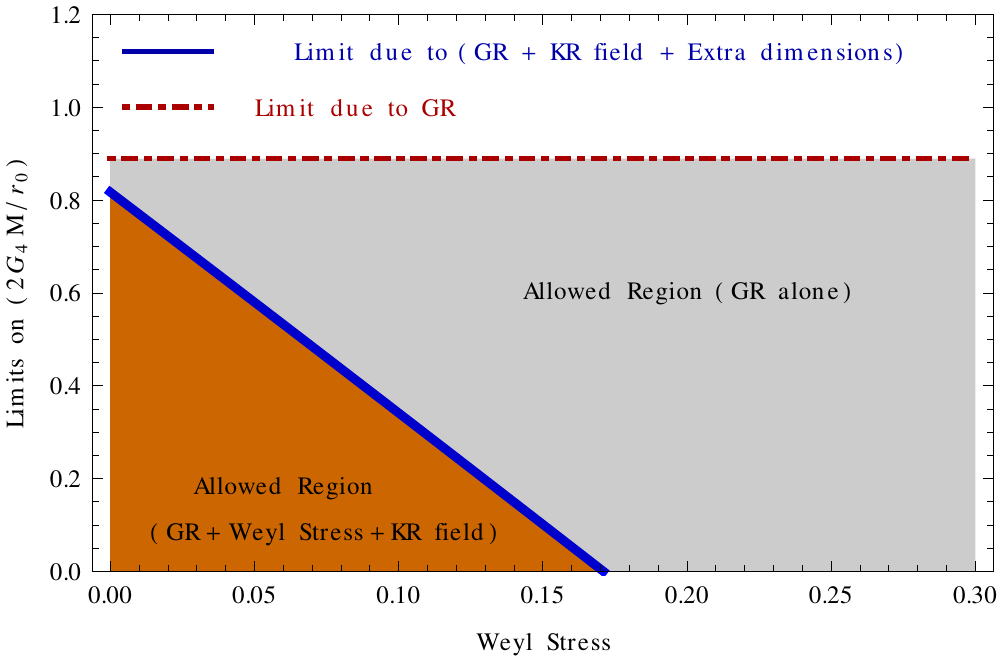}~~~
\includegraphics[scale=0.75]{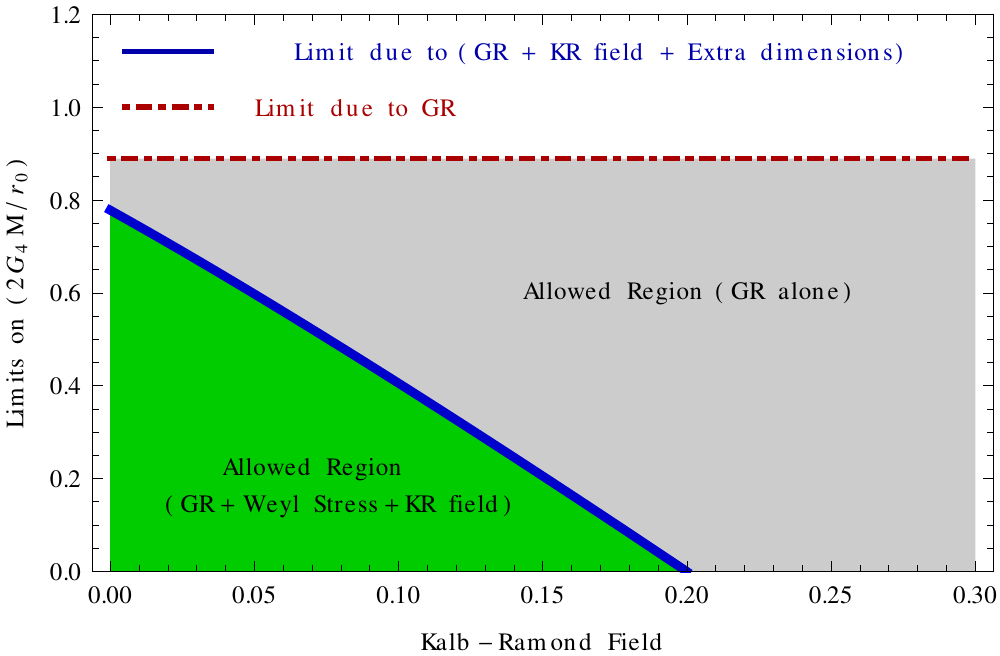}
\end{center}
\caption{Allowed region for $2G_{4}\mathcal{M}/r_{0}$ is being depicted as the \KR field and the dark radiation makes their appearance in the picture. It is clear that the size of the allowed region decreases as the \KR field strength as well as the Weyl stress inherited from the bulk increases. Thus one can add lesser amount of mass compared to \gr\ to the stellar structure, at a fixed radius, if the \KR field as well as higher dimensions are present. The first figure is drawn for fixed $\tilde{h}_{0}$ and $\rho_{0}$, while the other one is for fixed $P_{0}$ and $\rho_{0}$ respectively. See text for more discussions.}
\label{Fig_Buch_02}
\end{figure}
Regarding the integral, one can perform it by noting that the average density decreases outwards as we have elaborated earlier. In particular, from \ref{Eq_N_time} it follows that $m(r)/r>(m(r_{0})/r_{0}^{3})r^{2}$, making the above inequality stronger. Defining $M$ to stand for $m(r_{0})$ one can immediately integrate the above expression leading to,
\begin{align}
e^{-\lambda_{0}/2}-\frac{r_{0}^{2}}{1-e^{-\lambda _{0}}}\left[\frac{1}{2r_{0}^{2}}\left(1-e^{-\lambda_{0}}\right)+\frac{8\pi G_{4}}{2\lambda _{\rm T}}\frac{\rho_{0}^{2}}{2}
+\frac{2}{8\pi G_{4}\lambda _{\rm T}}\left(\frac{U_{0}}{2}+P_{0}\right)+\frac{9}{2}\sqrt{\frac{8\pi G_{4}}{\lambda _{\rm T}}}\tilde{h}_{0}^{2}\right]\left(1-e^{-\lambda_{0}/2}\right)>0~,
\end{align}
where again any quantity with subscript `0' ensures that it has been obtained on the surface of the star. Working out the above inequality in an explicit manner the corresponding limit on $\exp(-\lambda _{0})$ yields, 
\begin{align}\label{Eq_Buch_N_02}
1-e^{-\lambda _{0}}<\frac{4}{9}\Bigg[1-\frac{3}{2}\Big(\frac{8\pi G_{4}}{2\lambda _{\rm T}}\frac{\rho_{0}^{2}}{2}
&+\frac{2}{8\pi G_{4}\lambda _{\rm T}}\left(\frac{U_{0}}{2}+P_{0}\right)+\frac{9}{2}\sqrt{\frac{8\pi G_{4}}{\lambda _{\rm T}}}\tilde{h}_{0}^{2} \Big)r_{0}^{2}
\nonumber
\\
&+\sqrt{1+\frac{3}{2}\left(\frac{8\pi G_{4}}{2\lambda _{\rm T}}\frac{\rho_{0}^{2}}{2}
+\frac{2}{8\pi G_{4}\lambda _{\rm T}}\left(\frac{U_{0}}{2}+P_{0}\right)+\frac{9}{2}\sqrt{\frac{8\pi G_{4}}{\lambda _{\rm T}}}\tilde{h}_{0}^{2} \right)r_{0}^{2}} \Bigg]~.
\end{align}
Since the metric functions are continuous across the surface of the star, $e^{-\lambda _{0}}$ appearing in the above inequality can be replaced by the corresponding metric element in the exterior region in the limit $r\rightarrow r_{0}$. The exterior solution without the \KR field have been derived in \cite{Dadhich:2000am} and corresponds to, $U=-(P/2)=-(4/3)\pi G_{4}\lambda _{\rm T}(|q|/r^{4})$. Thus with the above choices for $U$ and $P$ but including the \KR field as well, leads to the following static and spherically symmetric solution, 
\begin{align}
e^{-\lambda}&=1-\frac{2G_{4}\mathcal{M}}{r}-\frac{3P_{0}r_{0}^{4}}{8\pi G_{4}\lambda _{\rm T}}\frac{1}{r^{2}}-\sqrt{\frac{8\pi G_{4}}{\lambda _{\rm T}}}\frac{h_{0}^{2}r_{0}^{4}}{r^{2}}
+\mathcal{O}(r^{-3});\qquad \tilde{h}(r)=\tilde{h}_{0}\left(\frac{r_{0}}{r}\right)^{2}+\mathcal{O}(r^{-3})~,
\nonumber
\\
e^{\nu}&=1-\frac{2G_{4}\mathcal{M}}{r}-\frac{3P_{0}r_{0}^{4}}{8\pi G_{4}\lambda _{\rm T}}\frac{1}{r^{2}}-5\sqrt{\frac{8\pi G_{4}}{\lambda _{\rm T}}}\frac{h_{0}^{2}r_{0}^{4}}{r^{2}}
+\mathcal{O}(r^{-3})~.
\end{align}
Here $P_{0}$ stands for the value of the ``dark pressure'' on the stellar surface which is positive definite and the relation $2U_{0}+P_{0}=0$ \cite{Dadhich:2000am,Germani:2001du} can be used to replace all the ``dark radiation'' term by ``dark pressure'' on the stellar surface. Further substituting the metric element $e^{-\lambda _{0}}$, evaluated at the surface of the star from the above equation in \ref{Eq_Buch_N_02} one finally obtains the following bound on the ADM mass of the stellar object,
\begin{align}\label{Buch_Eq_F}
\frac{2G_{4}\mathcal{M}}{r_{0}}<\frac{4}{9}\Bigg[1-\frac{3}{2}\Big(\frac{8\pi G_{4}}{2\lambda _{\rm T}}\frac{\rho_{0}^{2}}{2}
+\frac{3P_{0}}{4\pi G_{4}\lambda _{\rm T}}&+6\sqrt{\frac{8\pi G_{4}}{\lambda _{\rm T}}}\tilde{h}_{0}^{2} \Big)r_{0}^{2}
\nonumber
\\
&+\sqrt{1+\frac{3}{2}\left(\frac{8\pi G_{4}}{2\lambda _{\rm T}}\frac{\rho_{0}^{2}}{2}
+\frac{3P_{0}}{16\pi G_{4}\lambda _{\rm T}}+\frac{9}{2}\sqrt{\frac{8\pi G_{4}}{\lambda _{\rm T}}}\tilde{h}_{0}^{2} \right)r_{0}^{2}} \Bigg]~.
\end{align}
This completes our discussion regarding the derivation of Buchdahl's limit in the presence of \KR field and higher spatial dimensions. For completeness, let us briefly comment on the stability of the above solution. For this purpose, as in the previous scenario one needs to consider the scaler, vector and tensor perturbations. Since the background is still spherically symmetric, all these perturbations satisfy the master equation presented in \ref{pert_tor}. The time dependence will again be through the $e^{\pm i\omega t}$ term and the stability essentially corresponds to the positivity of the potential $V_{s}$ appearing in \ref{pert_tor}. In the present context both $e^{\nu}$ and $e^{-\lambda}$ behaves as the corresponding metric elements associated with the Reissner-Nordstr\"{o}m black hole with a negative $Q^{2}$, pertaining to the smallness of the \KR parameter $h_{0}$. One can immediately verify, given the metric elements, that the potential $V_{s}$ is necessarily positive 
irrespective of the value of $s$ \cite{Berti:2015itd,Pani:2013pma,Toshmatov:2016bsb}. Thus a preliminary analysis suggest that the solution considered above is indeed stable. However, in order to get the full picture and a concrete statement regarding stability, one must work through the black hole perturbation theory and the associated quasi-normal modes. We hope to address these issues elsewhere.

As evident from \ref{Buch_Eq_F}, the bound on ADM mass $2G_{4}\mathcal{M}/r_{0}$ is small compared to the \gr\ value $8/9$ (see \ref{Fig_Buch_02}). This is in complete contrast with the result obtained in the previous section, where the Buchdahl's limit was higher compared to the corresponding situation in \gr. Thus in this particular scenario the maximum mass that a compact stellar object can inherit at a fixed radius will be less compared to \gr. Thus it will be more difficult to probe this particular scenario, since if one observes a compact stellar object with a $\mathcal{M}/r_{0}$ ratio less than $4/9$, it can either correspond to \gr\ or the current scenario. It will be problematic to disentangle these two affects. 
\section{Discussions}\label{Buch_Tor_Conc}

We have explicitly demonstrated how the presence of \KR field (or, equivalently spacetime torsion) as well as that of extra dimensions modify the Buchdahl's limit. While pursuing the above we have used some general principles, e.g., matter density should decrease outwards in order to arrive at an inequality that depends only on the $g_{rr}$ component of the metric. This is possibly originating from the fact that, only the three curvature of a four dimensional spacetime encodes the gravity degrees of freedom. The above result also provides a new perspective on the Buchdahl's limit and possibly an universal upper bound on the $g_{rr}$ component. In particular, if we consider the \KR field in four spacetime dimensions, it follows that the (mass/radius) ratio is larger than $4/9$, the value pertaining to \gr. Thus at a certain radius one can introduce extra matter to the compact object. Hence if it is possible to detect a compact object with (mass/radius) ratio larger than the \gr\ prediction one can infer 
about the possible presence of the \KR field. On the other hand, when extra spatial dimensions are introduced, the effect of the bulk \KR field induced on the brane hypersurface leads to interesting features. For example, the effective energy momentum tensor on the brane violates the weak energy condition but does satisfy the strong energy condition. Similarly, there will be additional contributions to the gravitational field equations on the brane inherited from the presence of bulk spacetime. These two effects will result into modifications in the Buchdahl's limit associated with a compact stellar object. However unlike the scenario in four spacetime dimensions, as extra spatial dimensions are included the bound on (mass/radius) ratio decreases in comparison to \gr. This makes it difficult to explore possible observational avenues in this context regarding the presence of the \KR field as well as that of extra dimensions. 
\section*{Acknowledgements}

Research of S.C. is supported by SERB-NPDF grant (PDF/2016/001589) from SERB, Government of India. 
\bibliography{References_Buchdahl_KR}

\bibliographystyle{./utphys1}
\end{document}